\documentclass{jfm}

\usepackage{graphicx}
\usepackage{newtxtext}
\usepackage{newtxmath}
\usepackage{natbib}
\usepackage{hyperref}
\usepackage{bm}
\usepackage{graphicx}
\usepackage[dvipsnames]{xcolor}
\usepackage{cleveref}
\usepackage{etoolbox}
\usepackage{tikz}
\usepackage{ulem}
\usetikzlibrary{arrows.meta}
\usetikzlibrary{shapes.geometric}
\usepackage{pgfplots}
\usepackage{marginnote} 
\usepackage[inline]{enumitem}
\hypersetup{
    colorlinks = true,
    urlcolor   = blue,
    citecolor  = black,
}

\newcommand{\RomanNumeralCaps}[1]

\definecolor{lightblue}{RGB}{173,216,230} 

\title{Intermittency in turbulent emulsions}

\author{\aff{1}
  \corresp{\email{JFMEditorial@cambridge.org}},
  H.-C. Smith\aff{1}
 \and J.Q. Long\aff{2}}

\author{
M. Crialesi-Esposito\aff{1},
G. Boffetta\aff{2},
L. Brandt\aff{3,4},
S. Chibbaro\aff{5}
\and 
S. Musacchio\aff{2}
}

\affiliation{
\aff{1}  INFN, Sezione di Torino, via Pietro Giuria 1, 10125, Torino, Italy
\aff{2} Dipartimento di Fisica and INFN, Universit\`a degli Studi di Torino, via P. Giuria 1, 10125 Torino, Italy.
\aff{3} FLOW Centre, KTH Royal Institute of Technology, Stockholm, Sweden
\aff{4} Department of Energy and Process Engineering, Norwegian University of Science and Technology(NTNU), Trondheim, Norway
\aff{5} Universit\'e Paris-Saclay, CNRS, LISN, 91400 Orsay, France
}

\begin{document}
\maketitle

\begin{abstract}
We investigate the statistics of turbulence in emulsions of two-immiscible fluids of same density. We compute for the first time velocity increments between points conditioned to be located in the same phase or in different phases and examine their probability density functions (PDF) and the associated structure functions (SF). This enables us to demonstrate that the the presence of the interface reduces the skewness of the PDF at scales below the Kolmogorov-Hinze scale and therefore the magnitude of the energy flux towards the dissipative scales, which is quantified by the third-order SF. The analysis of the higher order SFs shows that multiphase turbulence is more intermittent than single-phase turbulence. In particular, the local scaling exponents of the SFs display a saturation about the Kolmogorov-Hinze scale and below, which indicates the presence of large velocity gradients across the interface. Interestingly, the statistics approach of classic homogeneous isotropic turbulence when significantly increasing the viscosity of the dispersed phase.
\end{abstract}



{\bf MSC Codes }  {\it(Optional)} Please enter your MSC Codes here

\section{Introduction}

Emulsions, \emph{i.e.} mixtures composed of two immiscible 
(totally or partially) liquids
with similar densities, are extremely common in industrial
applications environment such as pharmaceuticals
\citep{nielloud2000pharmaceutical, spernath2006microemulsions},
food processing \citep{mcclements2015food}
and oil production 
\citep{kokal2005crude,mandal2010characterization,kilpatrick2012water}. 
Emulsions are also important in geophysical applications: as example, when  oil or industrial wastes spill into water streams (from rivers to oceans),  the oil droplet
distribution becomes fundamental for quantifying the environmental damage 
\citep{Li1998,french2004oil,gopalan2010turbulent}.

At very low-volume-fraction, turbulent emulsions are mainly
characterized by the breakup of droplets.
The droplet size distribution is produced by the turbulent stresses
and the feedback of the dispersed phase on the carrier flow is small, and often neglected. 
The dynamics of droplet breakup, for a dilute emulsion in a homogeneous
and isotropic turbulent flow was initially investigated by \citet{Kolmogorov1949} and
\citet{Hinze1955}, who derived an expression for the maximum size of droplets resisting breakup
as a function of the flow characteristics and the fluid properties.
This is usually referred to as the Kolmogorov-Hinze (KH) scale.
Recent numerical investigations of droplets, bubbles and emulsions 
confirmed the general validity of the KH theory, both in isotropic and homogeneous
turbulence \citep{Perlekar2014,Mukherjee2019,Riviere2021,crialesi2022,girotto2022build,begemann2022effect},
and in anisotropic flows \citep{Soligo2019,Rosti2020,Pandey2022},
 while some theoretical corrections were lately proposed to account for scale-local nature of the process \citep{crialesi2022interaction,qi2022fragmentation}. 

At finite volume fraction of the dispersed phase, the distribution of the droplet sizes
results from the interplay between breakup and coalescence.
In this regime, the presence of droplets also modulates the underlying turbulence, affecting the flow statistics both at large 
\citep{Yi2021,wang2022turbulence} and small scales
\citep{Mukherjee2019,Freund2019,Vela2021,crialesi2022}.
In particular, the presence of a dispersed phase alters significantly the statistics at the small scales,
producing large deviations from the average values
of dissipation and vorticity  \citep{crialesi2022}.

In this work, we address the effects of the dispersed phase 
on the velocity increments of the turbulent flow at moderate ($10\%$)
and high ($50\%$) volume fractions. 
We study the role of the interface separating the two phases
by examining the statistics of velocity increments
between two points which are conditioned
to be either in the same phase or in different phases.
We find that the most important deviations from the statistics of
single-phase flows, quantified by the PDFs of the velocity increments,
are concentrated in regions around the interface,
i.e.\ when the two points belongs to different phases.
Moreover, we show that the amplitude of the third-order structure function (SF)
is reduced because of the contribution of the points located around the interface.
This is associated with a reduction of the flux
of kinetic energy in the turbulent cascade, which, in combination with the surface tension term, alters significantly the energy transport across scales. 
Finally, we discuss the effects of the droplets on the local
scaling exponents of the high-order structure functions,
which display a striking saturation at small scales. 

The remaining of this paper is organized as follows. 
In Section~\ref{sec2} we introduce the numerical
method adopted for the simulations, 
Section~\ref{sec3} is devoted to the presentation of the results
and Section~\ref{sec4} summarises the main conclusions. 

\section{Methodology}
\label{sec2}

We consider the velocity field ${\bm u}({\bm x},t)$ obeying
the Navier-Stokes equations
\begin{equation}
	\rho \left( \partial_t {\bm u} + {\bm u} \cdot {\bm \nabla u} \right) = 
	- {\bm \nabla} p + {\bm \nabla} \cdot \left[\mu \left({\bm \nabla u}
	+ {\bm \nabla u}^{T} \right) \right] + {\bm f}^{\sigma} + {\bm f}
	\label{eq2.1}
\end{equation}
and the incompressibility condition
${\bm \nabla} \cdot {\bm u}=0$. In \Cref{eq2.1} $p$ is the pressure and
$\rho$ is the density and $\mu({\bm x},t)$ is the local 
viscosity.
The surface tension force is represented by the term 
${\bm f}^{\sigma}=\sigma \xi \delta_S {\bm n}$ where $\sigma$ is the 
surface tension coefficient, $\xi$ is the local interface curvature, ${\bm n}$ the
surface normal unit vector and $\delta_S$ represents a delta function which 
ensures that the surface force is applied at the interface only
\citep{tryggvason2011direct}. The last term ${\bm f}$ is a constant in time
body force which sustains turbulence by injecting energy at large scales. 
Here, we adopt the so-called ABC forcing \citep{Mininni2006} which reads 
${\bm f}=(A\sin(k_f z)+C\cos(k_f y),B\sin(k_f x)+A\cos(k_f z),
C\sin(k_f y)+B\cos(k_f x))$.
The forcing scale is given by $L_f = 2\pi/k_f$. 

We solve \Cref{eq2.1} in a triply-periodic, cubic domanin of size $L=2\pi$,
discertized on a staggered uniform Cartesian grid. 
Spatial derivative are discretised with a second-order centered finite difference scheme
and time integration performed by means of a second-order Adam-Bashford scheme.
To reconstruct the interface, we use the algebraic Volume of Fluid method, MTHINC, introduced by \cite{Ii2012}.
A constant-coefficient Poisson equation is obtained using the
pressure splitting method \citep{Dodd2014}, which is solved
using a Fast Fourier Transform direct solver.
All the simulations have been performed with the code
FluTAS, described in \cite{crialesi2023flutas},
where further details on the numerical methods employed in this study can be found. 

We consider four different cases, all using a fixed ABC forcing 
with $A=B=C=1$ and $k_f=2 \pi/L_f=2$.
The reference single phase (SP) simulation assumes viscosity $\mu=0.006$, corresponding to a Taylor-scale Reynolds number
$Re_\lambda=15 k^2/(\nu\varepsilon)^{1/2}\approx137$, with
$\varepsilon=\nu\langle(\nabla{\bm u})^2\rangle$  the energy dissipation
rate, $k$ the turbulent kinetic energy and $\nu$ the kinematic viscosity. 
We vary the volume fraction $\alpha=V_d/V$,
defined as the ratio between the volume of the 
dispersed phase $V_d$ and the total volume $V=L^3$,
and the viscosity ratio $\gamma=\mu_d/\mu_c$.
As regards different volume fractions, we analyze the two cases with
$\alpha=0.1$ (hereafter MP10) and $\alpha=0.5$ (hereafter MP50), while keeping
$\gamma=1$ for both cases. Finally, we study the case $\alpha=0.1$ and
$\gamma=100$ (hereafter MPM). For all  multiphase (MP) simulations the density
ratio among the two phases is kept equal to $1$ and
the Weber number $We=\rho L_f u_{rms}^2/\sigma$=42.6. 

All the simulations are performed at a resolution $N=512$ which is
sufficient to resolve all the scales \citep[see][]{crialesi2022}; statistics are accumulated over several large eddy turnover times 
$T=L_f/u_{rms}$ once statistical stationary conditions have been reached.
For further details on the simulation setup, we refer the reader to \cite{crialesi2022}. 

\section{Statistics of the multiphase flow}
\label{sec3}

\begin{figure}
	\centering
	\includegraphics[width=0.7\textwidth]{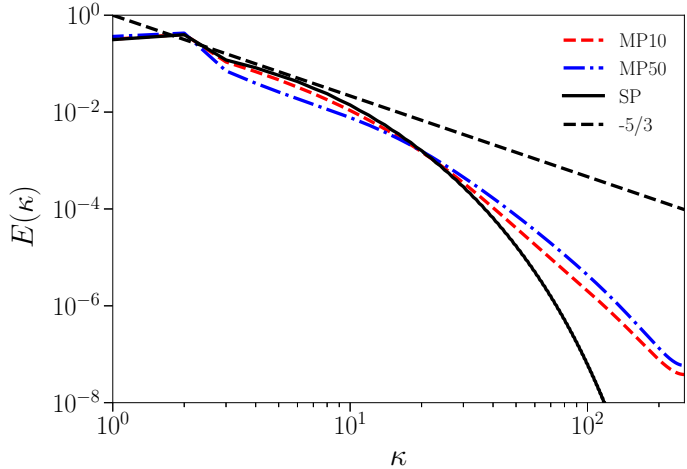}
	\caption{Kinetic energy spectra of SP flow (black continuous line),
	  and MP flows at $\alpha=0.1$ (red dashed line)
          and $\alpha=0.5$ (blue dash-dotted line)}
	\label{fig1}
\end{figure}

In turbulent multiphase flows, part of the kinetic energy of the carrier phase
is absorbed at large scales by the deformation
and breakup of the interface of the dispersed phase,
while the coalescence of small droplets, their surface oscillations
and relaxation from high local curvature re-inject energy
in the carrier phase at scales smaller
than the Kolmogorov-Hinze scale~\citep{crialesi2022interaction}.
The consequences of this complex exchange of energy between the two phases
are evident in the kinetic energy spectrum shown in \Cref{fig1}.
Comparing the spectra of a multiphase flow with that of a single-phase
flow sustained by the same forcing, we observe a suppression of energy 
at low wavenumbers (i.e. large scales) and an enhancement at high wavenumbers.  
This effect increases with the volume fraction $\alpha$ of the 
dispersed phase~\citep{Mukherjee2019,crialesi2022}. 

\begin{figure}
	\includegraphics[width=0.5\textwidth]{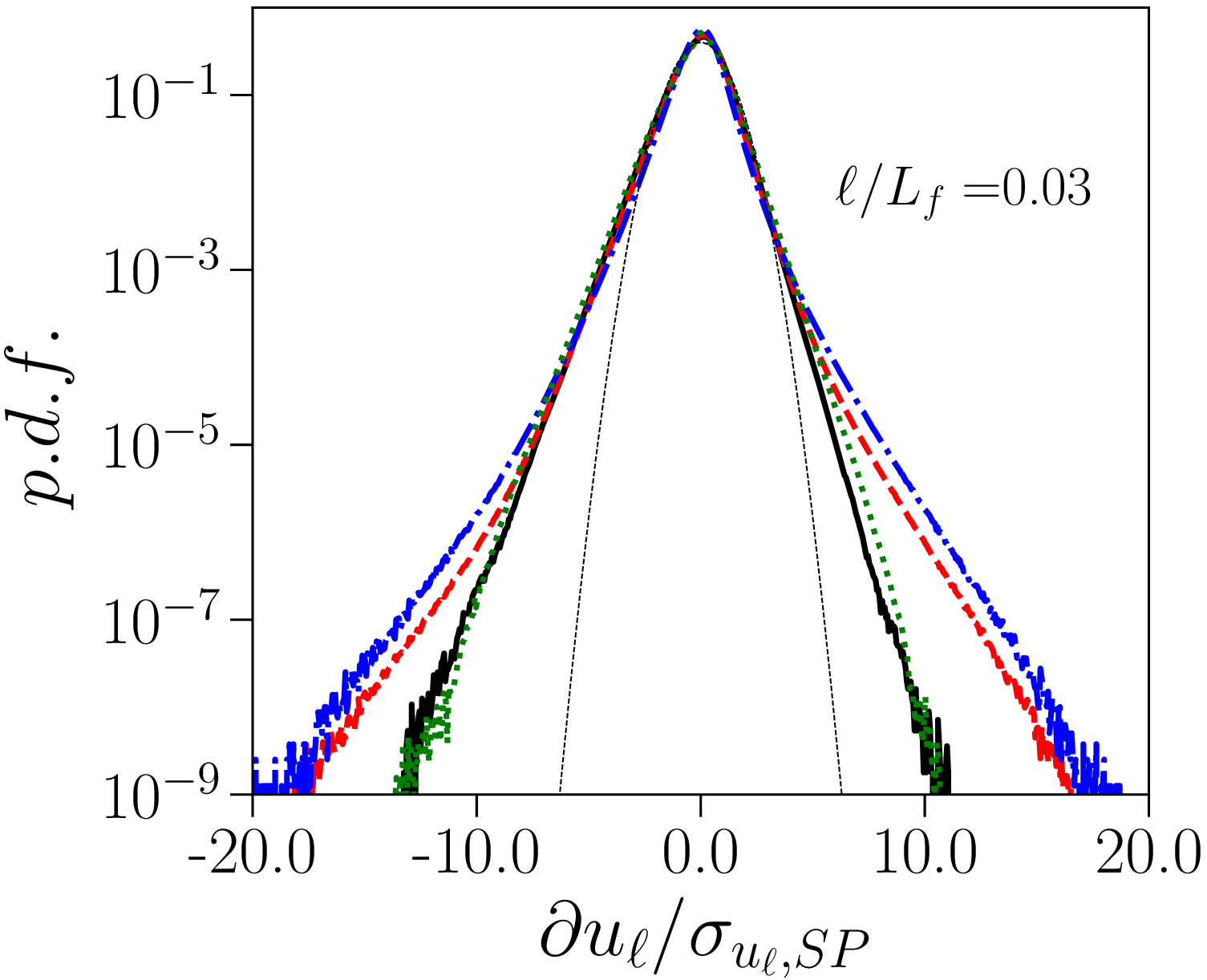}
	\put(-150,160){\large\textbf{(\textit{a})}}
	\includegraphics[width=0.5\textwidth]{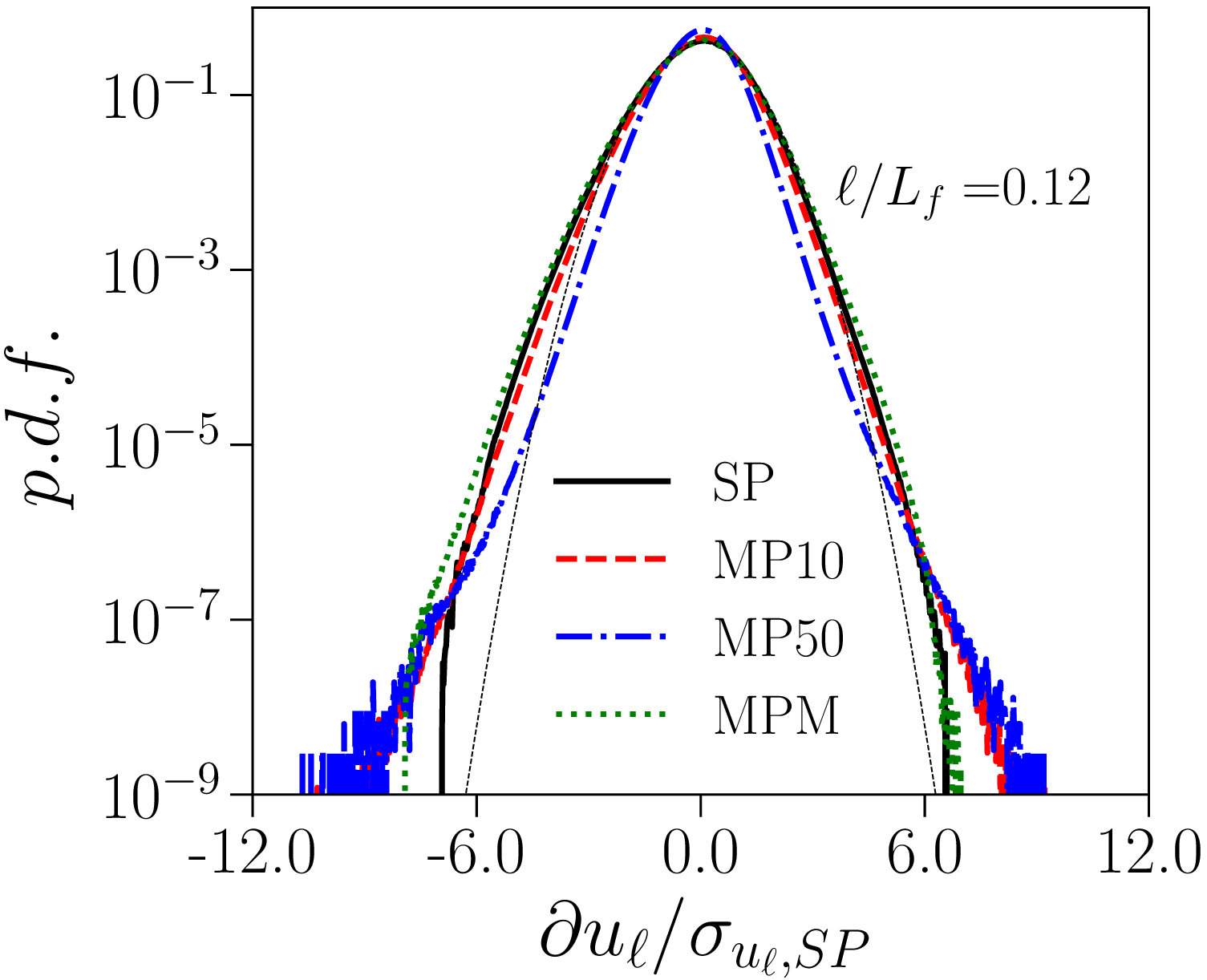}
	\put(-150,160){\large\textbf{(\textit{b})}}
	\caption{PDF of velocity increments at distance
          $\ell = 0.03L_f$ (left panel)
          and $\ell = 0.12L_f$ (right panel),
          normalized by the standard deviation of the SP case. 
          The single-phase Kolmogorov scale is at $\ell\approx0.008L_f$
}
	\label{fig2}
\end{figure}

Because of the injection of energy at small scales, due to the droplet
dynamics, we expect higher intermittency of the velocity
fluctuations in the MP flow than in the SP flow at fixed amplitude of the
external forcing. 
In order to quantify this effect we compute the probability density 
functions (PDF) of the longitudinal velocity increments 
$\delta_{\ell} u=({\bm u}({\bm x}_2)-{\bm u}({\bm x}_1)) \cdot ({\bm x}_2-{\bm x}_1)/\ell$
at distance $\ell=|{\bm x}_2-{\bm x}_1|$.
The comparison of the PDFs at two scales within the inertial range, shown in \Cref{fig2},
confirms that the velocity increments have larger fluctuations
in the case of MP flows, in particular at 
smaller values of $\ell$.
This effect increases with the concentration $\alpha$ of the dispersed phase.
We also observe that in the case $\gamma=100$ (i.e. when the dispersed
phase is much more viscous than the carrier phase) the 
effect of the droplets on the velocity increments vanishes
due to the damping of fluctuations in the dispersed phase,
and we recover the statistics of the SP flow. 

Note that the PDF shown in \Cref{fig2} are computed over the
full simulation domain, i.e. the velocity increments are computed
among points ${\bm x}_{1,2}$ which can belong to both phases unconditionally. 
To understand the role of the interface 
in the turbulent statistics, we therefore compute the PDF of the velocity 
increments conditioned to points belonging to the same or to different 
phases. Hence, we introduce three different PDFs of the velocity 
increments depending on which phase the two points ${\bm x}_1$ 
and ${\bm x}_2$ belong to. 
We denote by $P_{cc}$, $P_{dd}$ and $P_{cd}$ the PDFs relative to 
points belonging only to the carrier phase $c$, only to the dilute phase $d$ 
and to both phases, respectively. 
We remark that, for small values of $\alpha$,
the statistics in the two phases are different.
For $\alpha=0.5$ and $\gamma=1$
the two phases are equivalent and therefore $P_{dd}=P_{cc}$.

\begin{figure}
	\includegraphics[width=0.5\textwidth]{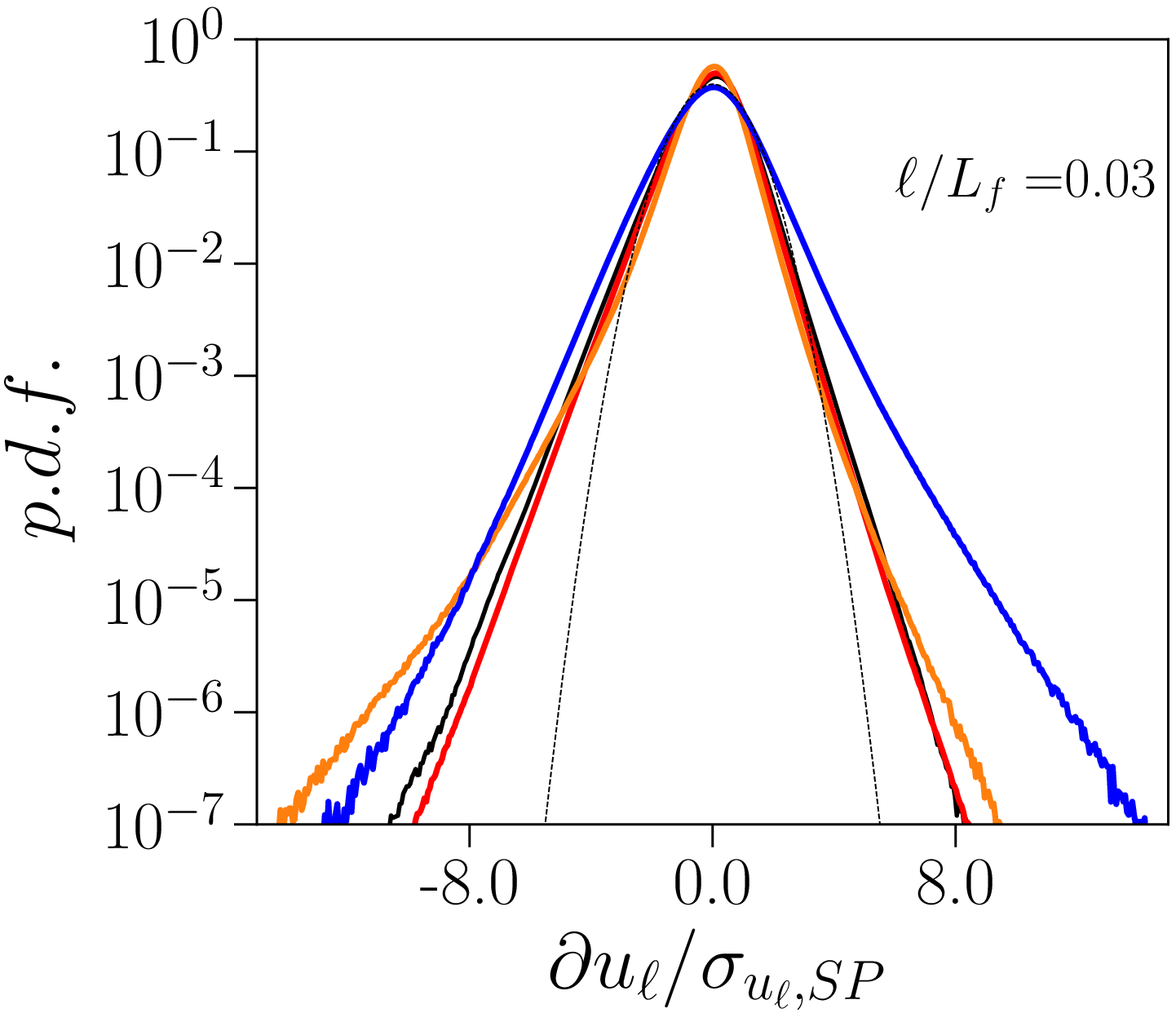}
		\put(-150,160){\large\textbf{(\textit{a})}}
	\includegraphics[width=0.5\textwidth]{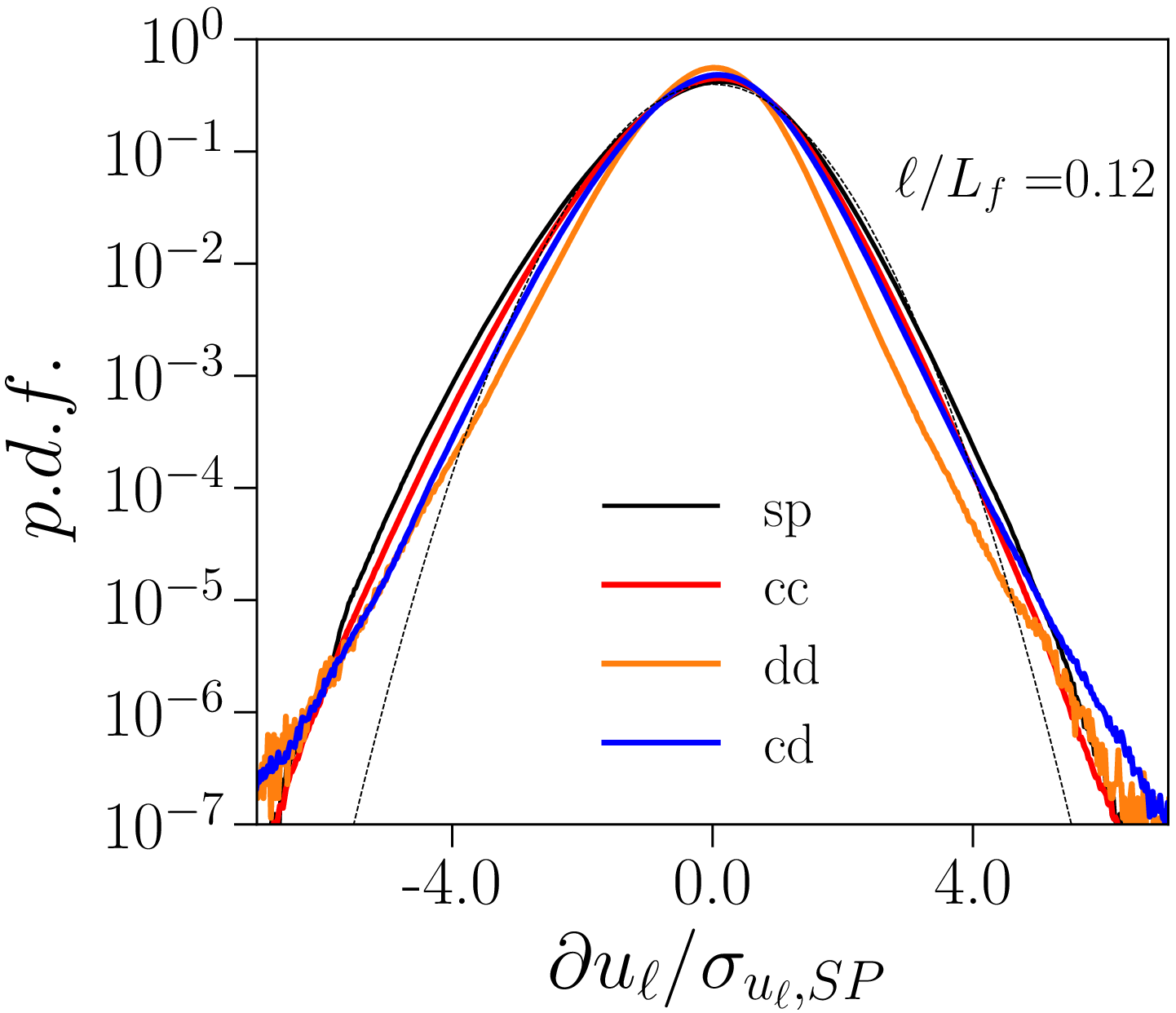}
		\put(-150,160){\large\textbf{(\textit{b})}}
	
	\includegraphics[width=0.5\textwidth]{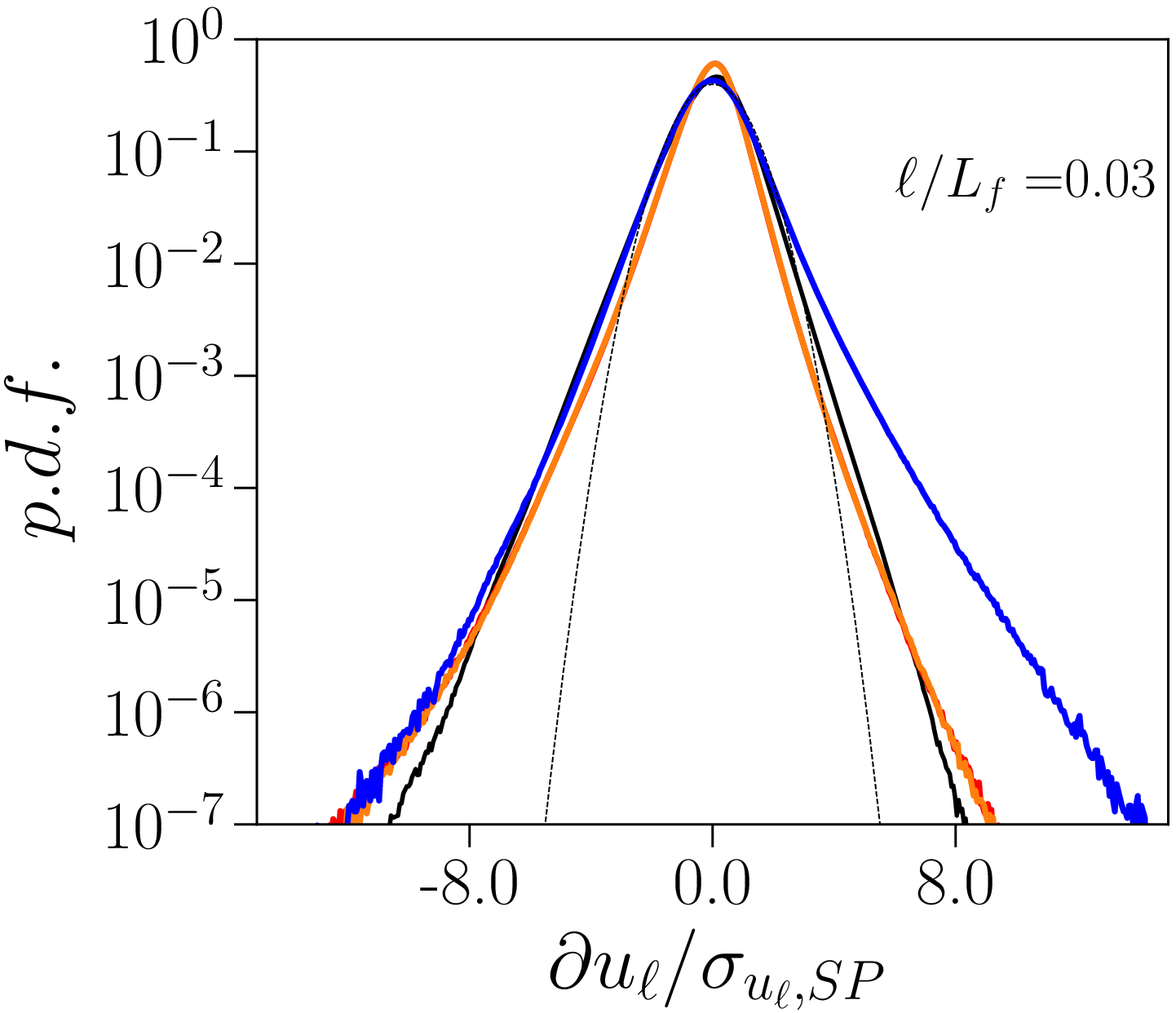}
		\put(-150,160){\large\textbf{(\textit{c})}}
	\includegraphics[width=0.5\textwidth]{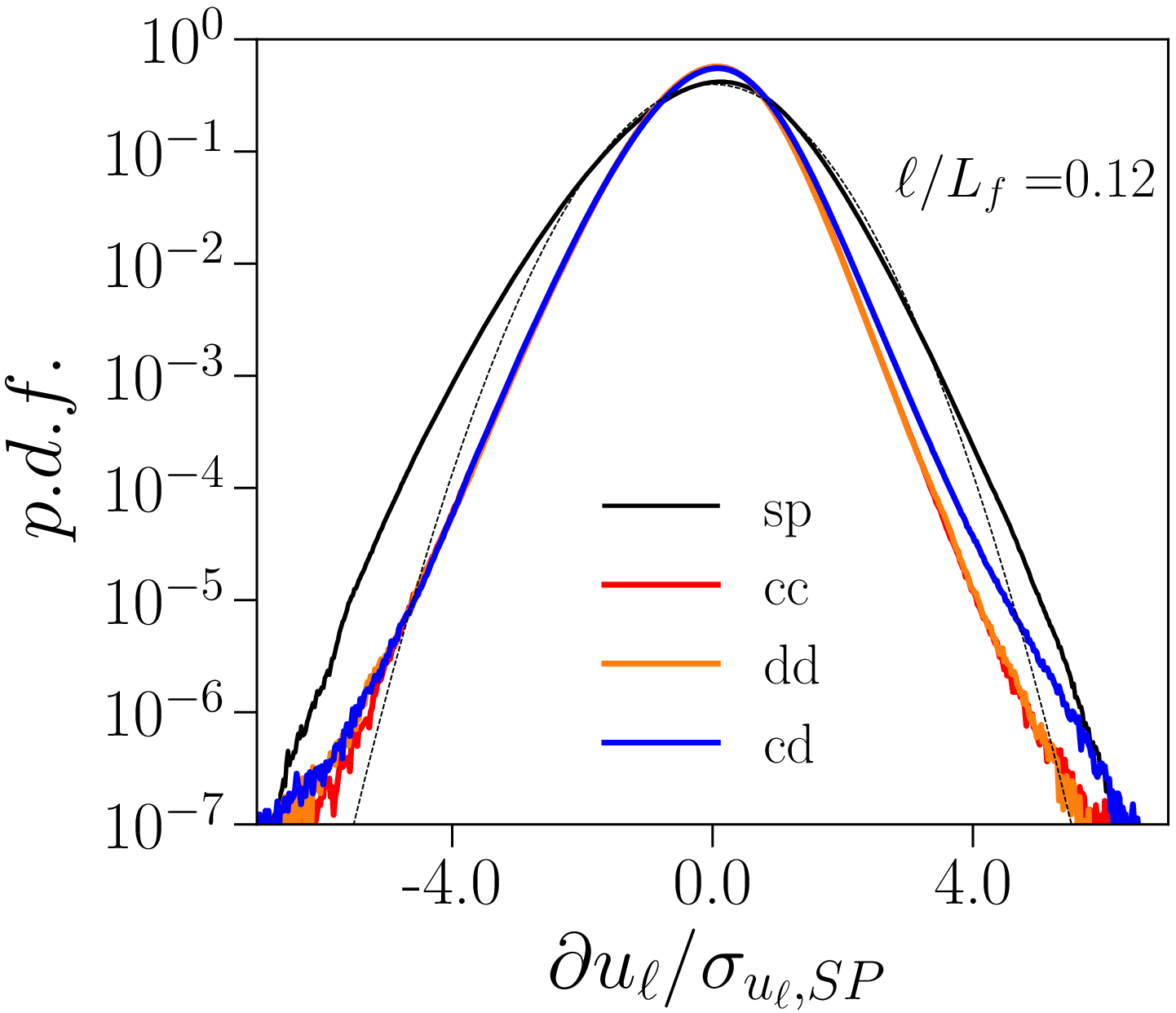}
		\put(-150,160){\large\textbf{(\textit{d})}}
	\caption{PDFs of velocity increments conditioned to the phases
		on which the two velocities are measured: $cc$ (both points in 
		the carrier phase, red line), $dd$ (both points in the dispersed
		phase, orange line), $cd$ (one point in each phase, blue line). 
		Upper panel: Simulation MP10 with $\alpha=0.1$ and $\gamma=1$.
                Lower panel: Simulation MP50 with $\alpha=0.5$ and $\gamma=1$.
		Black line: PDF of velocity increments for a single-phase simulation with 
		the same parameters of the MP simulation. 
		Dashed black line: Gaussian distribution.  
		All the PDFs are rescaled with the variance of the SP case.
	}
	\label{fig3}
\end{figure}

\Cref{fig3} (panels \textit{a}, \text{b})
shows the conditional PDF (normalized with the
corresponding variance of the SP case) pertaining the simulation with
volume fraction $\alpha=0.1$.
First, we note that the PDF of the carrier phase is not too far from that of the
SP case (and this is the case also for the variance). 
In the dispersed phase, on the contrary, velocity 
increments develop relatively larger tails at small separations. 
Remarkably, the PDF $P_{cd}$ develops 
the largest tails at small scale (panel \textit{a}), a clear indication of the role of
the interface for small-scale intermittency in MP flows.

Similar observations can be made for the emulsion with $\alpha=0.5$, 
shown in \Cref{fig3} (panels \textit{c}, \text{d}). 
As expected 
$P_{cc}=P_{dd}$, and also in this case the data show that
the leading contribution to the increased intermittency at small scales comes from velocity increments across the interface,
$P_{cd}$. Note that, although the 
shapes of $P_{cd}$ 
are similar for MP10 and MP50, their contribution to the overall flow statistics is different because
of the different statistical weight (i.e.\ the different extension of the total interface).  

A remarkable feature shown in \Cref{fig3}
is that the skewness of $P_{cd}$ at small scales is opposite 
(i.e. positive) to that of $P_{cc}$.
We remind that the sign of the skweness is linked to the direction
of the turbulent energy cascade
via the third-order velocity structure function 
(SF) defined as $S_3(\ell)=\langle (\delta_{\ell} u)^3 \rangle$.
In the case of SP flows,
under the assumption of statistical stationariety, homogeneity and isotropy,
the Kolmogorov $4/5$ law gives $S_3(\ell) = -(4/5) \varepsilon \ell$,
where the viscous energy dissipation rate $\varepsilon$
is equal to the flux of the turbulent cascade \citep{Frisch1995a}. 
The negative skewness of the PDF of the longitudinal velocity increments
is therefore related to the direction of the energy transfer 
and the negative amplitude of $S_3(\ell)$ is proportional to the energy flux. 

\begin{figure}
  	\includegraphics[width=0.5\textwidth]{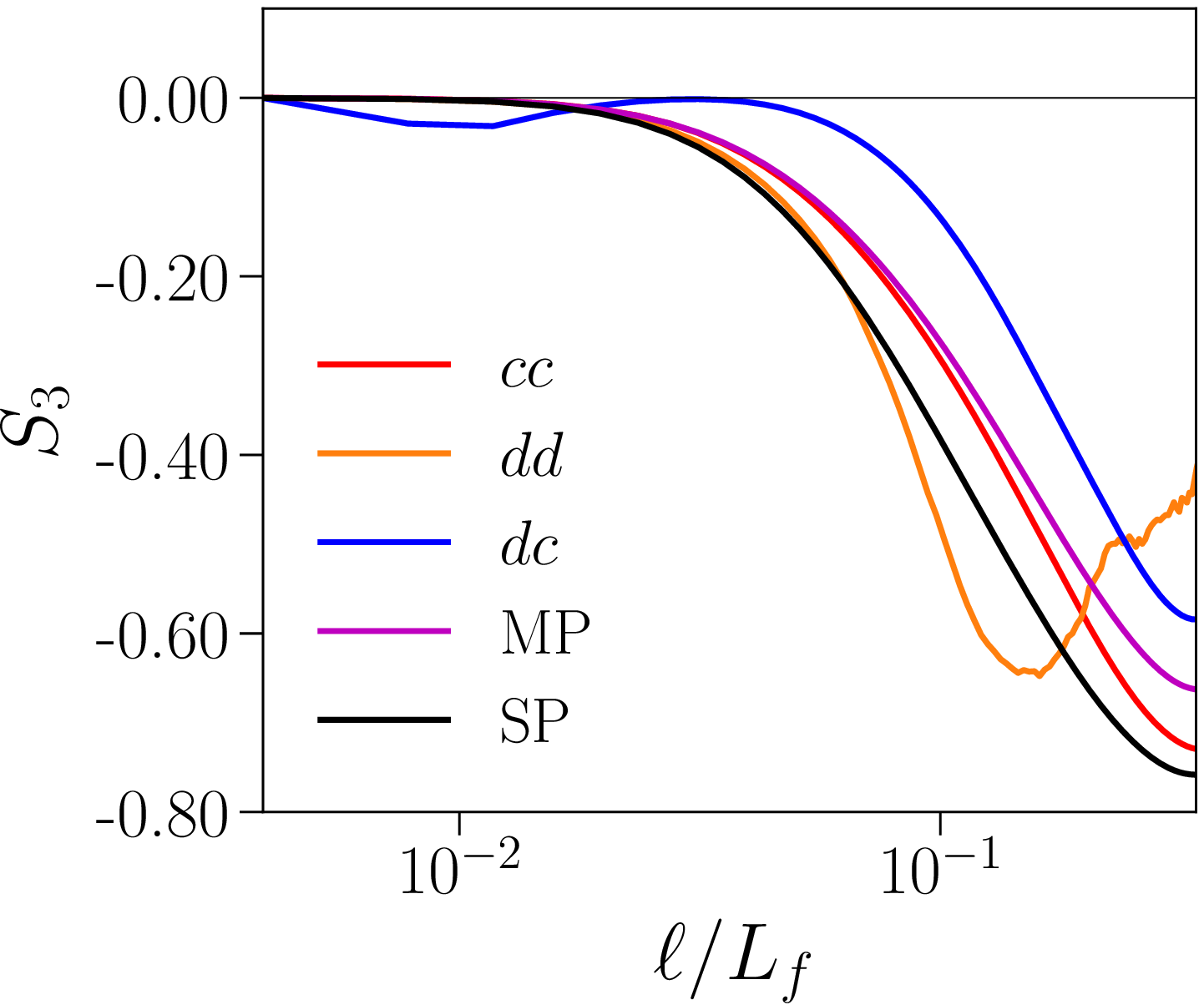}
  		\put(-150,160){\large\textbf{(\textit{a})}}
  		\includegraphics[width=0.5\textwidth]{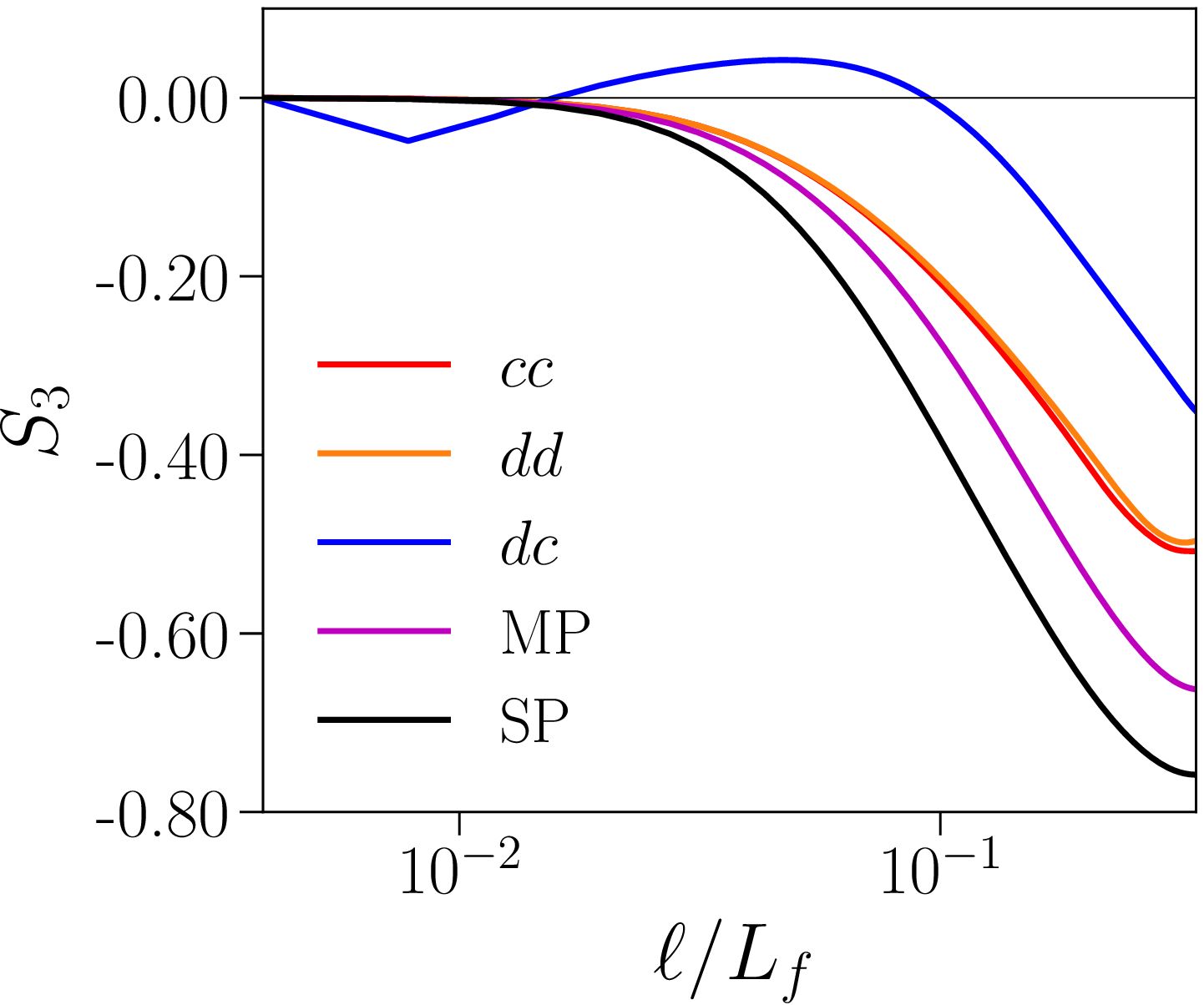}
  			\put(-150,160){\large\textbf{(\textit{b})}}
	\caption{Third-order stucture function $S_3(\ell)$ averaged on the
		whole domain (violet line) or conditioned on the 
		two phases of the flows. Simulation with
                $\alpha=0.1$ (left panel) 
		$\alpha=0.5$ (right panel) and $\gamma=1$.
                The black line represents the SP case.
         }
	\label{fig4}
\end{figure}

In MP flows, because of the opposite sign of the skewness of $P_{cd}$
with respect to $P_{cc}$,
we expect that the presence of the interface reduces the energy flux
associated to the turbulent cascade.
This can be quantified by looking at the
third-order velocity structure function
$S_3(\ell)=\langle (\delta_{\ell} u)^3 \rangle$, whose average can be unconditioned,
or conditioned to two points belonging
either to the carrier phase $S_3^{cc}(\ell)$,
or to the dispersed phase $S_3^{dd}(\ell)$,
or points located on different sides of the interface $S_3^{dc}(\ell)$. These are shown in \Cref{fig4}.

We see in the figure that the third-order SF of the MP turbulent flows 
is qualitatively similar to the
SP flow when averaged over the whole domain, yet with a smaller amplitude. 
This is due the fact that part of the turbulence energy is used to break 
the interface and the direct transfer of energy to small scales is reduced \citep{crialesi2022}.
If we consider the 
same quantity averaged over one of the two phases only:
$S_3^{cc}(\ell)$ and $S_3^{dd}(\ell)$, which are equivalent in a binary flow,
the magnitude of the flux increases and approaches the SP limit, indicating that the turbulent cascade is not significantly affected when considering flow structures living in one of the two phases.
On the contrary, the flux across two points belonging to different phases
is strongly suppressed: the associated $S_3^{dc}(\ell)$  
is closer to zero and even changes sign at intermediate scales for $\alpha=0.5$
(consistently with what suggested in
\Cref{fig3}).
The physical interpretation is that the interface
``decouples'' the velocity fields in the two phases which become
less correlated and therefore with a reduced energy flux,
signaled by the reduction of $S_3(\ell)$.
The precise behavior of $S_3^{dc}(\ell)$ depends on 
the value of $\alpha$, as shown by the comparison with the case $\alpha=0.1$
(see \Cref{fig4}, left panel). 
Positive values of  $S_3^{dc}(\ell)$, hint however to the possibility of scale-local
backscatter, which becomes more relevant at increasing volume fractions.
Nevertheless, the reduction of the energy flux at intermediate scales
is a general feature, independent on $\alpha$.

\begin{figure}
	\includegraphics[width=0.5\textwidth]{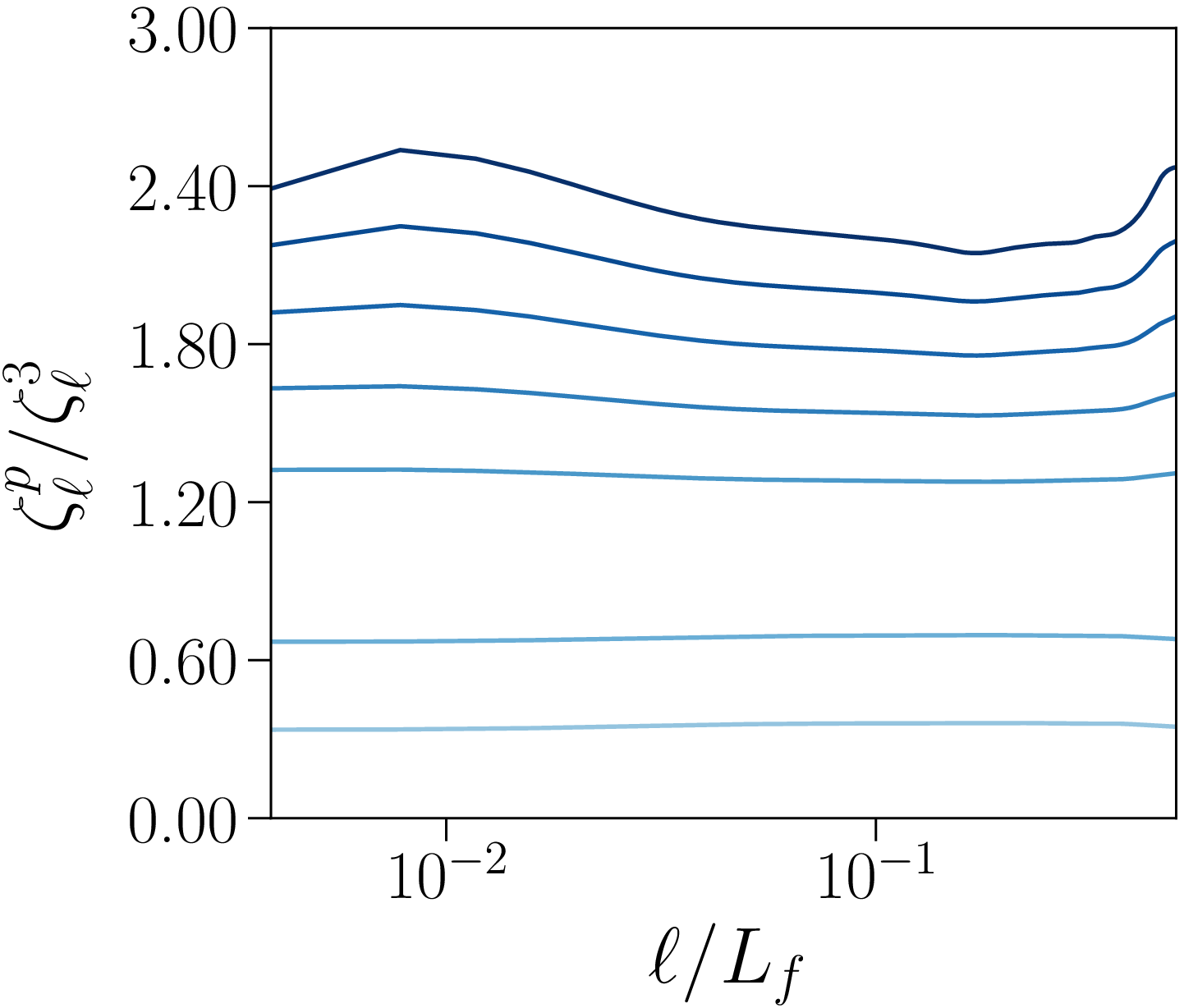}
		\put(-150,160){\large\textbf{(\textit{a})}}
	\includegraphics[width=0.5\textwidth]{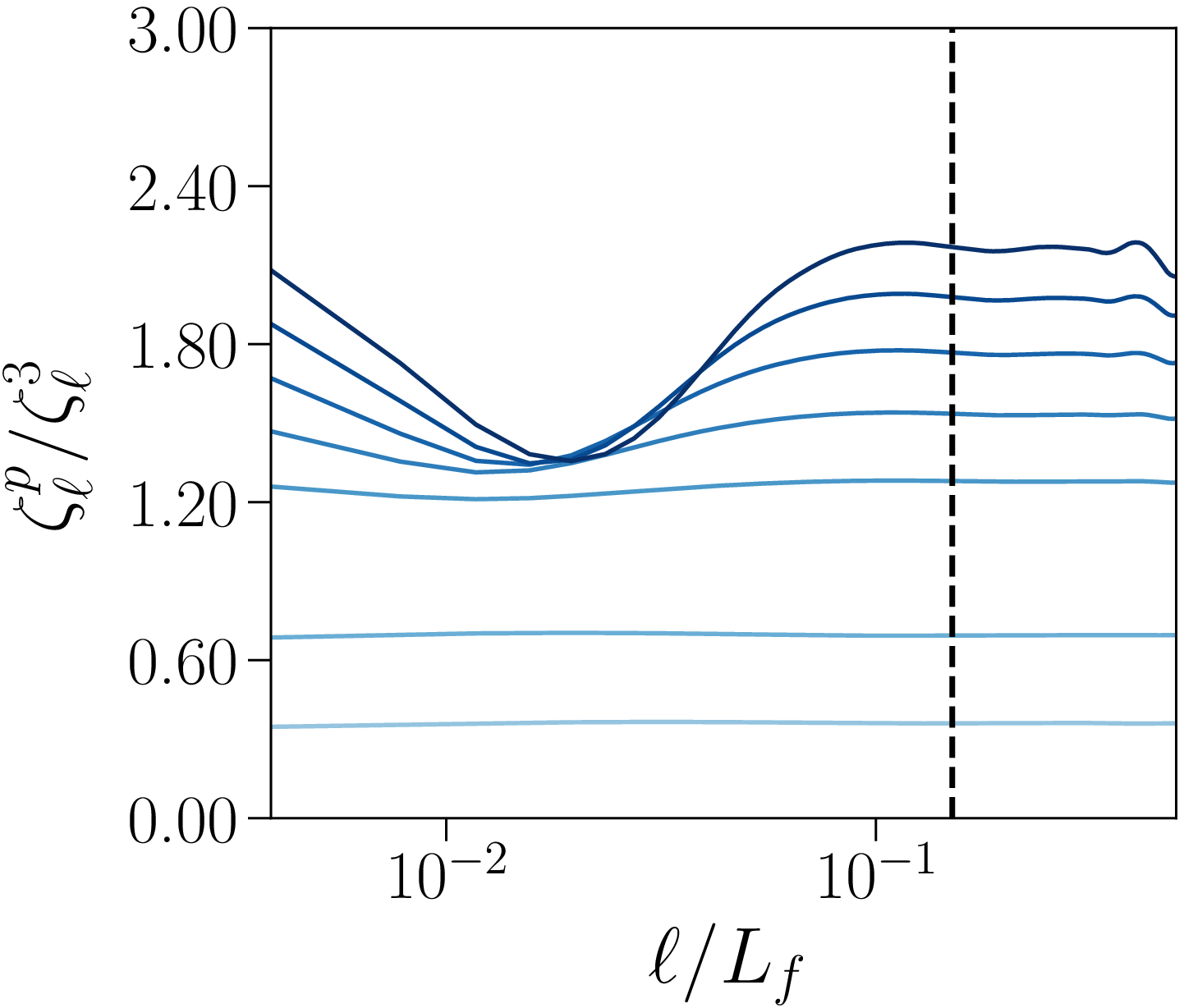}
		\put(-150,160){\large\textbf{(\textit{b})}}
		
	\includegraphics[width=0.5\textwidth]{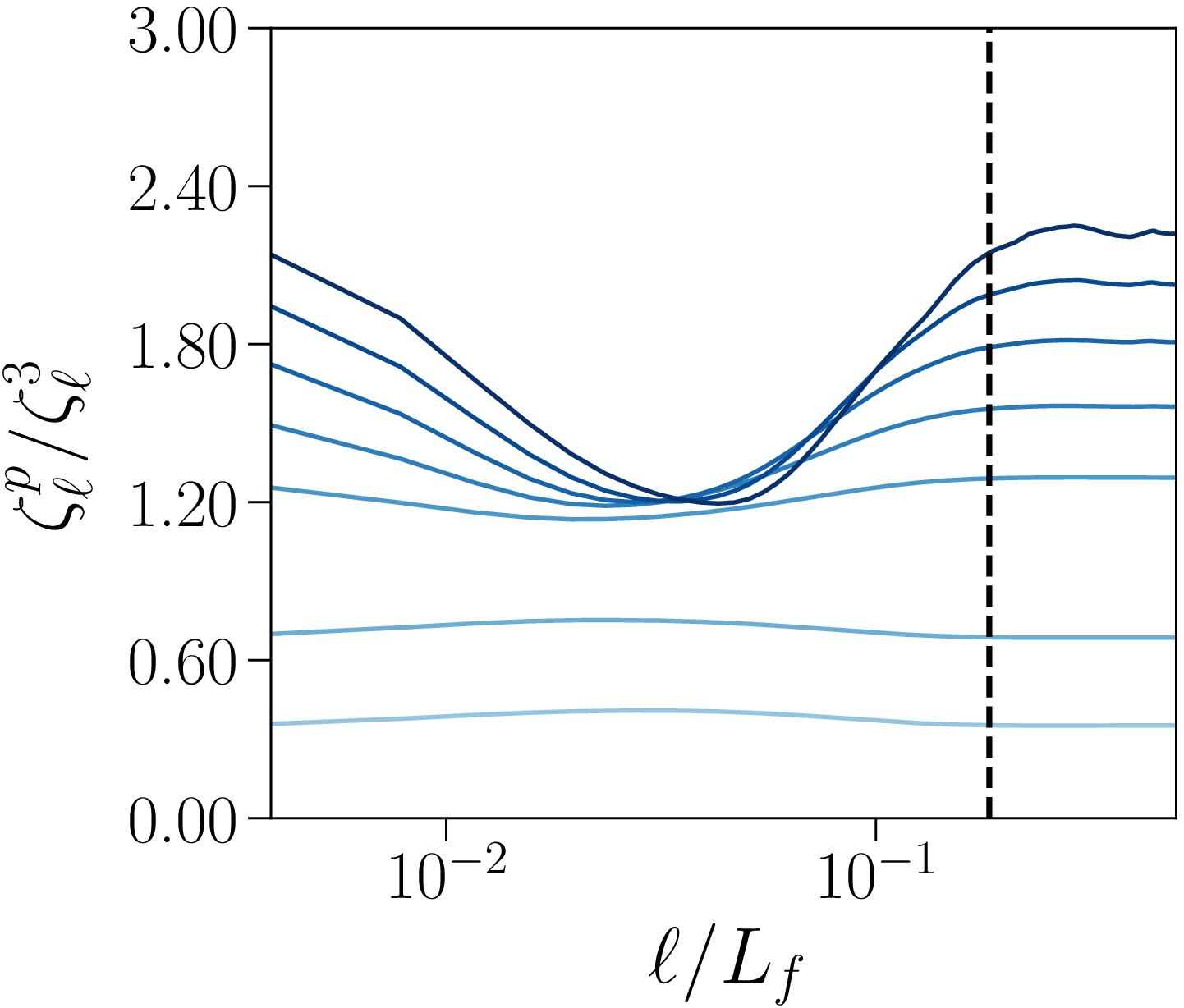}
		\put(-150,160){\large\textbf{(\textit{c})}}
	\includegraphics[width=0.5\textwidth]{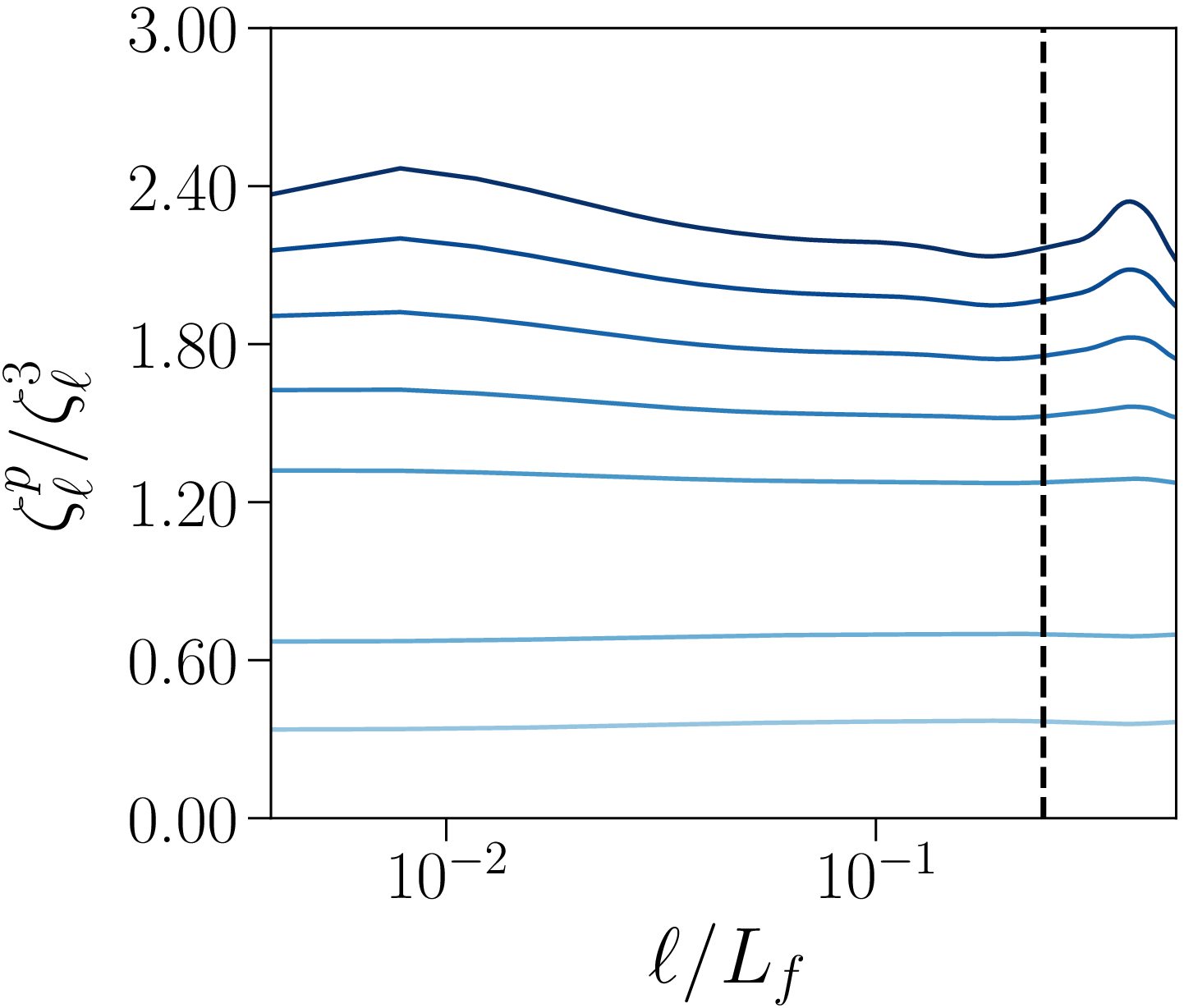}
		\put(-150,160){\large\textbf{(\textit{d})}}
	\caption{Local scaling exponents $\zeta_p$ of the structure functions,
          for the single phase (Upper left panel),
          case MP10 at $\alpha=0.1$ (Upper right panel),
          case MP50 at $\alpha=0.5$ (Lower left panel),
          case MPM at $\gamma=100$ (Lower right panel).
          In each figure, the exponents of different order $p$ assume 
          increasing values. 
          Vertical black dashed lines represent the Kolmogorov-Hinze scale, 
          computed in \cite{crialesi2022interaction}.
          Curves for $p=3$ is omitted.
         }  
	\label{fig5}
\end{figure}

The effects of the presence of a dispersed phase 
on the statistics of the velocity fluctuations affects also
the scaling behavior of the structure functions
of the absolute values of the longitudinal velocity increments 
defined as $S^a_p(\ell) = \langle |\delta_{\ell} u|^p \rangle$.
It is well known that in SP flows the SFs display a power-law behavior
$S^a_p(\ell) \sim \ell^{\zeta^p}$ at scales $\ell$ in the inertial range
\citep{Frisch1995a}. 
In this context, intermittency manifests in the non-linear behavior of the
scaling exponents: $\zeta^p \neq p/3$. 
In the MP flows, because of the different physical processes which occur
at scales larger and smaller than the Kolmogorov-Hinze scale (dominated by brak-up and coalescence respectively),
we expect to observe a more complex scaling behavior. 
To address this issue, we compute the {\it local scaling exponents}
defined as the logarithmic derivative of the SFs
$\zeta^p_{\ell} = {\text{d} \log(S^a_p(\ell))}/{\text{d} \log(\ell)}$
and here applied to multiphase flows for the first time. 

The local scaling exponents $\zeta^p_{\ell}$ are displayed 
for $p \le 8$ in figure~\ref{fig5}, where they are
divided by the reference scaling exponent $\zeta^3_{\ell}$ of the third 
order SF.
In the SP case, panel \textit{a}, we find that the ratios $\zeta^p_{\ell}/\zeta^3_{\ell}$
are almost constant in the inertial range $ 0.09\le \ell/L_f \le 0.32$. 
In the MP flows, the value of the exponents are a little smaller but comparable
to that of the SP case only
at large scales;
we observe a dramatic decrease of the scaling exponents at scales
$\ell$ smaller than the KH scale 
$L_{KH} \approx 0.14 L_f$ for the case MP10 (panel \textit{b})
and  $L_{KH} \approx 0.19 L_f$ for the case MP50 (panel \textit{b}) (see vertical line in figure). 
In particular, we observe a striking saturation of the
scaling exponents of the high-order SF with $p \ge 5$
at scales $\ell \simeq 0.02 L_f$ for the MP50 case and
$\ell \simeq 0.04 L_f$ for the MP10 case.
The saturation of the scaling exponents of the high order SFs
reveals the presence of strong velocity
differences across the interface between the two phases,
which originates from the pressure jump at the interface
caused by the surface tension forces. 

Note also that the saturation of the exponents is not observed  when the dispersed phase presents higher viscosity, case MPM in panel \textit{d}. This is consistent with the previous observations in \Cref{fig2}, showing that when velocity gradients across the interface are significantly reduced (\emph{e.g.} by higher viscosity) no exponent saturation is observed.

\section{Conclusions}
\label{sec4}

We have discussed intermittency and scaling exponent obtained from direct numerical simulations (DNS) 
of turbulent emulsions at moderate ($10\%$) and high ($50\%$) volume fractions and two different values of the viscosity contrast.
As observed in previous works  \citep{Perlekar2019,Pandey2020,crialesi2022}, 
the presence of a deformable interface 
increases the intermittency in the flow and the energy content at small scales,
when the surface tension offers an alternative path for energy transport across
scales.

By investigating the statistics of the velocity increments conditioned to
points belonging to a single phase or to different phases, we demonstrate that
the increased intermittency is mostly due to the presence of strong velocity
differences across the interface between the carrier and the dispersed phase.

We also show that the presence of the dispersed phase causes a decrease of the
negative skewness of the PDF of the longitudinal velocity increments.  This is
associated with a reduction of the flux of the kinetic energy from the forcing
scale to the viscous scales. 
In other words, the presence of a deformable
interface affects the vortex stretching and tilting associated to the classic
turbulent energy cascade of single-phase flows.

This effect becomes remarkable at the highest volume fraction considered here,
when the flux related to points lying on either side of the interface gives a
positive contribution to the distribution skewness.
This suggests
a not-negligible  backscatter in multiphase flows, 
expected in proximity of the interface separating the two fluids. 
We interpret this reduced flux as due to the absorption and dissipation of part
of the kinetic energy of the turbulent flow by the deformation and break-up of
drops of the dispersed phase.

Finally, to understand the local properties of turbulence, we have analysed the longitudinal Structure Functions at higher orders. 
Interestingly,
at scales larger than the Kolmogorov-Hinze, the exponents are only slightly smaller than in the single-phase flow, 
which implies increased intermittency, yet a similar anomalous scaling.
More importantly, we report a neat saturation of the exponents for structure functions higher than 3 at scales smaller than the Kolmogorov-Hinze length. 
 This is typically related to a strongly intermittent dynamics and to the presence of jumps, here due to the pressure differences across the interface induced by the surface tension.

A further demonstration that the interface is responsible of the increased
intermittency is given by the results for the flow at viscosity ratio
$\gamma=100$. In this case, small-scale fluctuations are damped, 
especially
in the more viscous dispersed phase, and the statistics approach those of the
single-phase turbulence with no exponent saturation.

These observations may prove fundamental for understanding small scale dynamics
in multiphase flows and for their future sub-grid modelling. Indeed, our
results indicate that a correct model would need to account for the reduction
of the energy fluxes near an interface. Moreover, we have shown that the
turbulence statistics approach those of the single-phase flow when the droplets
consist of a highly viscous fluid. This suggests that, despite several global
measures seem to indicate a similar dynamics
\citep{Olivieri2022,yousefi2022transport}, the turbulence modulation is
significantly different in the case of rigid particles and deformable
intrusions.

\bibliographystyle{jfm}
\bibliography{references}


\backsection[Acknowledgements]{MCE, GB and SM acknowledge the support from the Departments of Excellence grant (MIUR) and INFN22-FieldTurb. 
The authors acknowledge computer time provided by the National Infrastructure for High
Performance Computing and Data Storage in Norway (Sigma2, project no. NN9561K) and by SNIC (Swedish
National Infrastructure for Computing).}

\backsection[Funding]{LB acknowledge the support from the Swedish Research Council via the multidisciplinary research environment INTERFACE, Hybrid multiscalemodelling of transport phenomena for energy efficient processes, Grant no. 2016-06119.  }

\backsection[Declaration of interests]{The authors report no conflict of interest.}

\backsection[Data availability statement]{Data are available from the corresponding author upon reasonable request.}



\end{document}